\documentstyle{article}

\begin{document}
\begin{flushright}
NUP-A-97-23
\end{flushright}
\vspace{1cm}
\begin{center}
{\Large
The Excess of HERA High$-Q^2$ Events and\\
\vspace{12pt}
 Leptoquarks in a Left-Right Symmetric Preon Model }

\end{center}
\vspace{2cm}
\begin{center}
{\Large
Motoo {\sc Sekiguchi}\footnote{motoo@phys.cst.nihon-u.ac.jp}, Hiroaki
{\sc Wada} and Shin {\sc Ishida}\\}

\vspace{1cm}
\end{center}
\begin{center}
{\Large
{\it
Atomic Energy Research Institute,\\
 College of Science and
Technology,\\
 Nihon University, Tokyo 101\\}}

\end{center}
{\large
\vspace{1cm}
\begin{center}
{\bf Abstract}
\end{center}
\vspace{12pt}
~~ An interpretation that the HERA excess events are due to intermediate
production and decay of composite leptoquarks in the left-right
symmetric preon model is 
given. Because of the preon-line rule, expected to be valid well, the
event-ratio of neutral to charged current interactions is predicted to
be one.}
\vspace{3cm}

\newpage

[{\it Introduction}]
Recently an interesting experimental fact which might reflect an opening
of new physics was reported:\cite{1,2} The deep-inelastic $e^{+} P$
scattering experiments with $\sqrt{s} = 300 GeV$ at HERA has observed an
excess of events, with the square of transferred momentum $Q^{2} > 15000
GeV^{2}$, over the expectation by the standard model(SM).
More recently the H1 and ZEUS experiments reported\cite{3} their new
results of neutral current and charged current events, corresponding to
the combined luminosity of 57.2 $pb^{-1}$: the H1 observed 8 events of
the neutral current with invariant mass $M = \sqrt{xs}$ ($x$ being the
Bjorken scaling variable) between 187.5 $GeV$ and 212.5 $GeV$ and $y =
Q^{2}/M^{2} > 0.4$, where $1.5 \pm 0.3$ events are expected by the SM.
The ZEUS observed 5 events of neutral current with $x > 0.55$ and $y >
0.25$, where $1.51 \pm 0.13$ events are expected.  Thus the H1 and ZEUS
experiments have observed totally the excess of 9 events of neutral
current. They  have also observed in the region with $Q^{2} > 10^{4}
GeV$ 28 events  of the charged current where $17.7 \pm 4.3$ events are
expected, accordingly the excess of 10 events of charged current 
interactions.

Immediately there have appeared many theoretical and phenomenological
interpretations of this excess. One of the most popular interpretations
seems to be due to a direct-channel production of leptoquarks, squarks
which couple with $e^{\pm}q$ in a $R$-symmetry-breaking supersymmetric
interaction.\cite{4} On the other hand we propose an interpretation of
the excess to be due to production of ^^ ^^ composite" leptoquarks which
are expected naturally to exist in a left-right symmetric preon model.
Some attempts\cite{5} from this line have been already done. Our
interpretation is simpler and more intuitive, and accordingly leads to
more clear-cut results, as a result of assuming the preon-line
rule,\cite{6} like the OZI-rule in the case of QCD, concerning the
hypercolor confining force of preons. As will be discussed shortly, this
line rule may be expected to be valid more rigorously in the relevant
case than in the case of QCD.

[{\it Preon model and leptoquarks}]
Following the many pioneering works of preon models,\cite{7,8} we take a
view point that the standard model is a low energy effective theory
among composite quarks, leptons and weak bosons which are consisting of
preons and/or anti-preons confined by a hypercolor
$SU(N)_{HC}$ gauge interaction.\cite{6,9} As basic preons we set up
$$
F=(F^{U}, F^{D}),  B^{i}=(C^{(i)}, S^{(i)}),
\eqno{(1)}
$$
where $F$'s are Dirac spinors belonging to an isodoublet of the global
$SU(2)$ and $B^{(i)}$'s are scalars carrying the generation
number(i=1,2,3). Both of $F$ and $B$ belong to the fundamental
$\underline{N}$-representation of $SU(N)_{HC}$ group. The quantum
numbers of preons are shown in Table I. 
\begin{table}[t]
\caption{Quantum numbers of preons }
\begin{center}
\begin{tabular}{l|c|c|c|c}
\hline\hline
   & $F^{U}$ & $F^{D}$ & $C^{(i)}$ & $S^{(i)}$ \\
\hline
hypercolor & $\underline{N}$ & $\underline{N}$ &
 $\underline{N}$ & $\underline{N}$ \\
color & $\underline{1}$ & $\underline{1}$ &
 $\underline{3^{*}}$ & $\underline{1}$ \\
charge & $1/2$ & $-1/2$ & $-1/6$ & $1/2$ \\
 B & $0$ & $0$ & $1/3$ & $0$ \\
 L & $0$ & $0$ & $0$ & $1$ \\
\hline
\end{tabular}
\end{center}
\end{table}
Here it is to be noted that only scalar preons $C$'s and $S$'s have the
baryon(or quark) and lepton numbers B and L, respectively.(The $C$'s
have also the color freedom, of which indices are omitted in this work).
 The hypercolor singet quarks  and leptons are supposed to have the
configurations\footnote{In our scheme the extra isosinglet bosons
$W^{(0)}_{\mu,h} =\overline{F}_{h}\gamma_{\mu}F_{h}$($h = L$ and $R$)
are also expcted naturally to exist.\cite{6} The lower limit of mass
values of these extra-bosons $\vec{W}_{R}$ and $W^{0}_{L}$ are studied
in our previous works.\cite{9}} as
$$
{u \choose d} = 
{F^{U}\overline{C}^{(1)} \choose F^{D}\overline{C}^{(1)}},  
{c \choose s} = 
{F^{U}\overline{C}^{(2)} \choose F^{D}\overline{C}^{(2)}}, 
{t \choose b} = 
{F^{U}\overline{C}^{(3)} \choose F^{D}\overline{C}^{(3)}},\\
\eqno{(2a)}
$$
$$
{\nu_{e} \choose e} = 
{F^{U}\overline{S}^{(1)} \choose F^{D}\overline{S}^{(1)}},  
{\nu_{\mu} \choose \mu} = 
{F^{U}\overline{S}^{(2)} \choose F^{D}\overline{S}^{(2)}},  
{\nu_{\tau} \choose \tau} = 
{F^{U}\overline{S}^{(3)} \choose F^{D}\overline{S}^{(3)}},  
\eqno{(2b)}
$$
In our scheme the gauge symmetry is naturally extended to that of
$SU(2)_{L} \times SU(2)_{R}$, and the respective hypercolor singlet 
gauge bosons $W_{L}$ and $W_{R}$ have the configurations as
$$
{ \vec{W}_{\mu,L}} \approx \left(\overline{F}_{L}{
\vec{\tau}}\gamma_{\mu} F_{L} \right), 
{ \vec{W}_{\mu,R}} \approx \left(\overline{F}_{R}{
\vec{\tau}}\gamma_{\mu} F_{R} \right),
\eqno{(3)}
$$
where $F_{L}$ and $F_{R}$ being the left- and right-handed component of
$F$. We also expect the existence of the other hypercolor singlet
particles in our scheme. In Table II these particles, having the
simplest configuration of two-body preon system, are listed. 
\begin{table}[t]
\caption{Hypercolor singlet composite bosons and respective production
channels }
\begin{center}
\begin{tabular}{cc}
\hline\hline
$\Phi^{(ij)}_{SC} \approx (S^{(i)}\overline{C}^{(j)})$
& $e^{+}P$ \\
$\Phi^{(ij)}_{SS} \approx (S^{(i)}\overline{S}^{(j)})$
& $e^{+}e^{-}$ \\
$\Phi^{(ij)}_{CC} \approx (C^{(i)}\overline{C}^{(j)})$
& $P\overline{P}$ \\
\hline
\end{tabular}
\end{center}
\end{table}

Out of the composite bosons in Table II the bosons of configuration
$$
\Phi^{(ij)}_{SC}\approx(S^{(i)}\overline{C}^{(j)}),
\eqno{(4)}
$$
having $B=-1/3$ and $L=1$, are our relevant leptoquarks.
In this work we assume that they are scalars, that is, to be the S-wave
bound-states of preons.

[{\it Line rule and properties of leptoquarks}]
As a confining force to combine preons we have supposed the $SU(N)_{HC}$
gauge interaction, and, furthermore, assume an approximate validity of
the preon-line rule\cite{6} like the OZI-rule in the case of QCD. The
OZI-rule is generally believed to be valid in the large-$N$
limit.\cite{10} In the preceding work of composite weak bosons in our
preon model it is shown that the number $N_{HC}$ is expected to be very
large.\cite{9,11} Accordingly we can expect that the preon-line rule in
the present case may be valid the more  rigorously than the OZI-rule.

The preon-line rule implies the conservations of respective
preon-numbers, which include the baryon, lepton and
generation\footnote{However, as for the conservation of generation
number see the remark on the mass eigenstates of leptoquarks given
later.} number conservations. Accordingly in our view-point there is no
problem of rapid proton-decay from the beginning. 
 The most-simple effective Lagrangian, satisfying the above requirement,
of Yukawa interaction of the leptoquarks with quarks and leptons is give
as
$$
{\cal L}=\sum_{i,j}\lambda_{SC}^{(ij)}\left(
   \overline{\nu}^{(i)}\Phi_{SC}^{(ij)}u^{(j)}
   +\overline{u}^{(i)}\Phi_{SC}^{(ij)\dagger}\nu^{(j)}
   +\overline{e}^{(i)}\Phi_{SC}^{(ij)}d^{(j)}
   +\overline{d}^{(i)}\Phi_{SC}^{(ij)\dagger}e^{(j)}\right),
\eqno{(5)}
$$
where $\nu^{(i)}$  and $e^{(i)}$($u^{(i)}$ and $d^{(i)}$ )are the i-th
generation leptons (quarks) given  in Eq(2). In Fig.1 we have shown the
corresponding Feymman diagrams to the Lagrangian (5), representing the
production(decay) process of the leptoquarks(anti-leptoquarks).In Table
III we have summarized the properties of our leptoquarks derived from
Eq.(5).

\begin{table}[t]
\caption{Production or Decay Ratio of $\Phi^{(ij)}_{SC}$ }
\begin{center}
\begin{tabular}{c|c|c}
\hline\hline
leptoquarks & prod. or decay channels & ratio \\
\hline
$\Phi^{(ij)}_{SC}$ & $(\overline{\nu}^{(i)}  u^{(j)})  /
(\overline{e}^{(i)}  d^{(j)}) $& 1 / 1\\
\hline
$\Phi^{(ij)\dagger}_{SC} $ & ($\overline{u}^{(i)}  \nu^{(j)}) / 
(\overline{d}^{(i)}  e^{(j)}) $ & 1 / 1 \\
\hline
\end{tabular}
\end{center}
\end{table}

[{\it HERA high $Q^{2}$ events and leptoqurks}]
The HERA excess events of neutral and charged current interactions by
the $e^{+}P$  collision experiment are interpreted to be due to the
intermediate $s$-channel production and decay of the leptoquarks
$\Phi^{(11)}_{S\bar{C}}$ as shown,respectively in
Fig 2 and in Fig 3, which are produced by the collision of the initial
$e^{+}$ and valence $d$-quark inside of the initial parton(see the
discussion given shortly).

The event ratio of neutral to charged event interactions is predicted to
be 1 as is given in Table III. Correspondngly the branching ratio
becomes
$$
\frac{\Gamma(\Phi^{(11)}_{SC} \longrightarrow e^{+}d)}{
\Gamma(\Phi^{(11)}_{SC} \longrightarrow all)}
=
\frac{\Gamma(\Phi^{(11)}_{SC} \longrightarrow \overline{\nu}^{+}u)}{
\Gamma(\Phi^{(11)}_{SC} \longrightarrow all)}
=\frac{1}{2}.
\eqno{(6)}
$$

It is interesting and encouraging that this prediction seems to be
consistent with the experimental results presently, which are mentioned
in the introduction.

 Here it is to be noted that our relevant leptoquarks are not able to be
produced by the high energy collision processes of
$e^{+}e^{-}$,$e^{-}P$, $P\bar{P}$ and $P\bar{P}$. It is also to be noted
that there are no contributions even through the exchange of leptoqurks
in the low energy scattering of these systems.

Now we go into some details by assuming
$$
\lambda_{SC}^{(ij)}=\lambda_{SC}\hspace{24pt}  \mbox{independent of
$(ij)$}
\eqno{(7)}
$$
on the Yukawa coupling constants in Eq.(5).

The production cross section of $\Phi^{(ij)}_{SC}$ is given, in the
leading order of $\alpha_{s}$ and in the narrow-width approximation for
$\Phi^{(ij)}_{SC}$ by 
$$
\sigma(e^{+}q \rightarrow \Phi_{SC}^{(ij)})
=
\frac{ \pi}{2S}\lambda_{SC}^{2}q\left({m_{\Phi}^2}/{s} \right),
\eqno{(8)}
$$
where $q(m_{\Phi}/s)$ is the $q$-parton distribution function\cite{12}
inside of $P$ at the $x=m_{\Phi}/s=0.44$ ($\sqrt{s}=300GeV$ being the
total C.M. energy of $e^{+}P$ system, $m_{\Phi}=200GeV$ obtained by the
H1 experiment\cite{1}). By virtue of Eq.(8) we can estimate the
production ratios for $\Phi^{(ij)}_{SC}$'s through the initial
valence-quarks and/or the sea-quarks as
$$
\frac{ \sigma_{sea}(e^{+}s \rightarrow \Phi_{SC}^{(12)})}
{\sigma_{val+sea}(e^{+}d \rightarrow \Phi_{SC}^{(11)})} \approx 0.02,
\frac{ \sigma_{sea}(e^{+}b \rightarrow \Phi_{SC}^{(13)})}
{\sigma_{val+sea}(e^{+}d \rightarrow \Phi_{SC}^{(11)})} \approx 0.01.
\eqno{(9)}
$$
Thus we have confirmed our conjecture that the HERA excess events are
due to production of leptoquarks $\Phi^{(11)}_{SC}$ through the
collision of $e^{+}$ and  $d$-quarks inside of proton. The value of
coupling constant $\lambda$ is determined to be
$$
\lambda_{SC} = 0.025
\eqno{(10)}
$$
by using the Eq.(8), where we have used the value of $\sigma$ estimated
as
$$
\sigma = \frac{19 \ \mbox{events}}{57 \mbox{pb}^{-1}};
 19\ \mbox{events} = 9 +10
\eqno{(11)}
$$
In Eq.(11) we have used the number of excess events N = 9(neutral
current) + 10(charged current) corresponding to ^^ ^^ total" luminosity
57 $\approx$ 23.7 + 37.5 $pb^{-1}$, as is mentioned in Introduction, and
$q_{d}(0.44)=0.083$.\cite{12} Using the value of $\lambda_{SC}$ in
Eq.(10) the total decay width of $\lambda^{11}$ is determined to be 
$$
\Gamma(\lambda_{SC}) = \frac{\lambda_{SC}^{2}}{4\pi} m_{\Phi}
= 10 MeV,
\eqno{(12)}
$$
which seems to guarantee the narrow width approximation made 
in obtaining Eq.(8).
However, this  value of $\Gamma$ given in Eq.(12) should be taken only
as a qualitative one, because there are given no considerations on the
possible contributions from the final hadrons with wee energy, which are
required to  exist in view of color symmetry.

Here we also add a comment on the number of possible leptoquarks: In the
recent report\cite{3} they also mentioned that the invariant mass
distributions of the the excess events of neutral current in the H1
experiment are clustering around $M = 200 GeV$, while in the ZEUS
experiment the excess events are clustering around $M = 220
GeV$\cite{3}, and accordingly that these excesses are unlikely to be due
to a single narrow resonance.
In our scheme, since all the scalar preons $C^{(i)}$($S^{i}$) of
quarks(leptons) have the same hypercolor charge, regardless of the
generation number $i$, the mass-eigen states of leptoquarks are
generally expected to be superposition of $S^{(i)}\overline{C}^{(j)}$
and the number of possible leptoquarks is, in
principle, 9 (aside from the color-freedom). It is to be noted that our
prediction of the branching ratio $1/1$ of neutral to charged current
events is still valid in this case, as is seen from Table III.
We might conjecture that the excess of neutral current events by the H1
and ZEUS experiments, seeming not due to a single resonance, are due to 
the two independent linear combinations of $\Phi_{SC}^{(11)}$ and
$\Phi_{SC}^{(12)}$ (or $\Phi_{SC}^{13}$). If this is the case, the line
rule concering the generation number is not valid.

[{\it Concluding remarks}]
In this letter we have given an interpretation of the HERA excess events
in the framework of the left-right symmetric preon model. We supose that
the excess events are due  to the intermediate s-channel production and
decay of leptoquarks, which are hypercolor singlet composite states of
the two-body preon system and are naturally to exist in the preon model.
Due to the preon-line rule, which is expect to be valid well, the event
ratio of neutral to charge current interactions is predicted to be 1. In
our view point the excess events due to the  relevant leptoquarks
$\Phi_{sc}^{(11)}$($\Phi_{sc}^{(11)})$ dose occur only with the initial
$e^{+}$$d$($e^{-}$$\bar{d})$ channel, that is,
$e^{+}$$N$($e^{-}$$\overline{N}$) channel. Finally we should like to
note that the excess events of similar nature due to intermediate
production and decay of the other hypercolor singlet boson,
$\Phi_{SS}$($\Phi_{CC}$) given in Table II, are also expected to occur
with the initial $e^{+}e^{-}$($P\overline{P}$) channel in the LEP II or
JLC (TEVATRON) experiment.

We would like to thank Dr. K. Tokushuku for valuable informations and
useful discussions. We are grateful to Professor T. Hirose for
encouragement. We thank Dr. K. Yamada, 
Dr. N. Honzawa, Mr. M-Y. Ishida and Professor M. Oda for continual
encouragement and useful discussions.

\newpage
\begin{center}
FIGURES
\end{center}

\begin{center}
Fig.1 {The production(decay) process of the
leptoquarks(anti-leptoquarks)}
\end{center}

\begin{center}
Fig.2. {Production and decay of the leptoquarks $\Phi_{SC}^{(11)}$
(neutral current)}
\end{center}

\begin{center}
Fig.3. {Production and decay of the leptoquarks $\Phi_{SC}^{(11)}$
(charged current)}
\end{center}

\end{document}